\documentclass[a4paper,fleqn,usenatbib]{mnras}

\usepackage[T1]{fontenc}
\usepackage{ae,aecompl}

\usepackage{graphicx}	
\usepackage{amsmath}	
\usepackage{amssymb}	
\usepackage{multirow}

\newcommand{\maxi}{\textit{MAXI}}
\newcommand{\nicer}{\textit{NICER}}

\newcommand{\vsgr}{V4641~Sgr}
\newcommand{\vcyg}{V404~Cyg}
\newcommand{\eighteen}{Swift~J1858.6--0814}
\newcommand{\exo}{EXO~0748$-$676}
\newcommand{\seight}{\eighteen}

\title[Dips and eclipses in Swift J1858.6$-$0814]{Dips and eclipses in the X-ray binary Swift~J1858.6--0814 observed with \nicer}

\author[D. J. K. Buisson et al.]{D. J. K. Buisson$^{1}$,\thanks{Email: d.j.k.buisson@soton.ac.uk}
  D. Altamirano$^{1}$,
  M. Armas Padilla$^{2,3}$,
  Z. Arzoumanian$^{4}$,  
\newauthor  
  P. Bult$^{5,4}$,
  N. Castro Segura$^{1}$,
  P. A. Charles$^{1}$,
  N. Degenaar$^{6}$,
  M. D{\'i}az Trigo$^{7}$,
\newauthor
  J. van den Eijnden$^{8,6}$,
  F. Fogantini$^{9,10}$,
  P. Gandhi$^{1}$,
  K. Gendreau$^{4}$,
  J. Hare$^{4}$,\thanks{NASA Postdoctoral Fellow}
\newauthor
  J. Homan$^{11,12}$,
  C. Knigge$^{1}$,
  C. Malacaria$^{13,14}$,\thanks{NASA Postdoctoral Fellow}
  M. Mendez$^{15}$,
  T. Mu\~noz Darias$^{2,3}$,
\newauthor
  M. Ng$^{16}$,
  M. \"{O}zbey Arabac\i$^{1,17}$,
  R. Remillard$^{16}$,
  T. E. Strohmayer$^{18}$,
\newauthor
  F. Tombesi$^{4,5,19,20}$,
  J. A. Tomsick$^{21}$,
  F. Vincentelli$^{1}$
  and D. J. Walton$^{22}$\\
  Affiliations are listed at the end of the paper.
}

\date{Accepted 2021 Mar 22. Received 2021 Mar 22; in original form 2021 Feb 26}

\pubyear{2020}

\begin{document}
\label{firstpage}
\pagerange{\pageref{firstpage}--\pageref{lastpage}}
\maketitle

\begin{abstract}
We present the discovery of eclipses in the X-ray light curves of the X-ray binary \eighteen.
From these, we find an orbital period of $P=76841.3_{-1.4}^{+1.3}$\,s ($\approx21.3$\,hours) and an eclipse duration of $t_{\rm ec}=4098_{-18}^{+17}$\,s ($\approx1.14$\,hours). We also find several absorption dips during the pre-eclipse phase.
From the eclipse duration to orbital period ratio, the inclination of the binary orbit is constrained to $i>70^\circ$. The most likely range for the companion mass suggests that the inclination is likely to be closer to this value than $90^\circ$.
The eclipses are also consistent with earlier data, in which strong variability (`flares') and the long orbital period prevent clear detection of the period or eclipses.
We also find that the bright flares occurred preferentially in the post-eclipse phase of the orbit, likely due to increased thickness at the disc-accretion stream interface preventing flares being visible during the pre-eclipse phase. This supports the notion that variable obscuration is responsible for the unusually strong variability in \eighteen.
\end{abstract}

\begin{figure*}
\includegraphics[width=\textwidth]{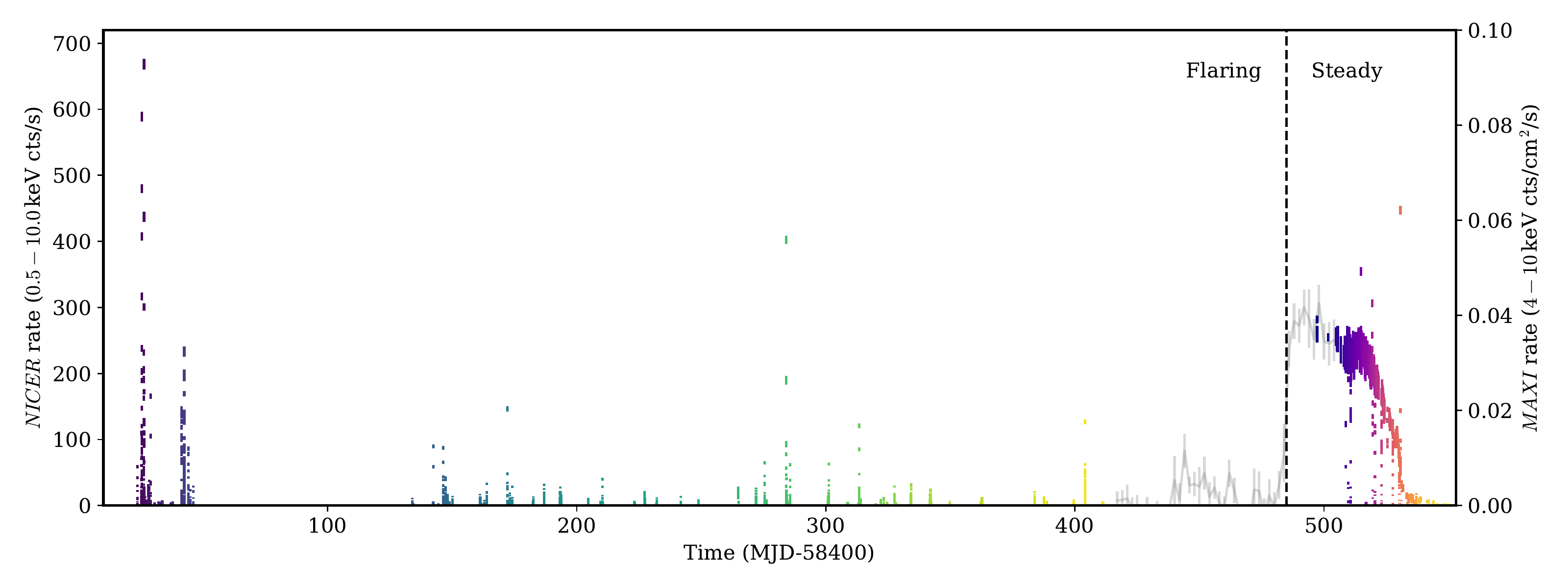}

\includegraphics[width=0.497\textwidth]{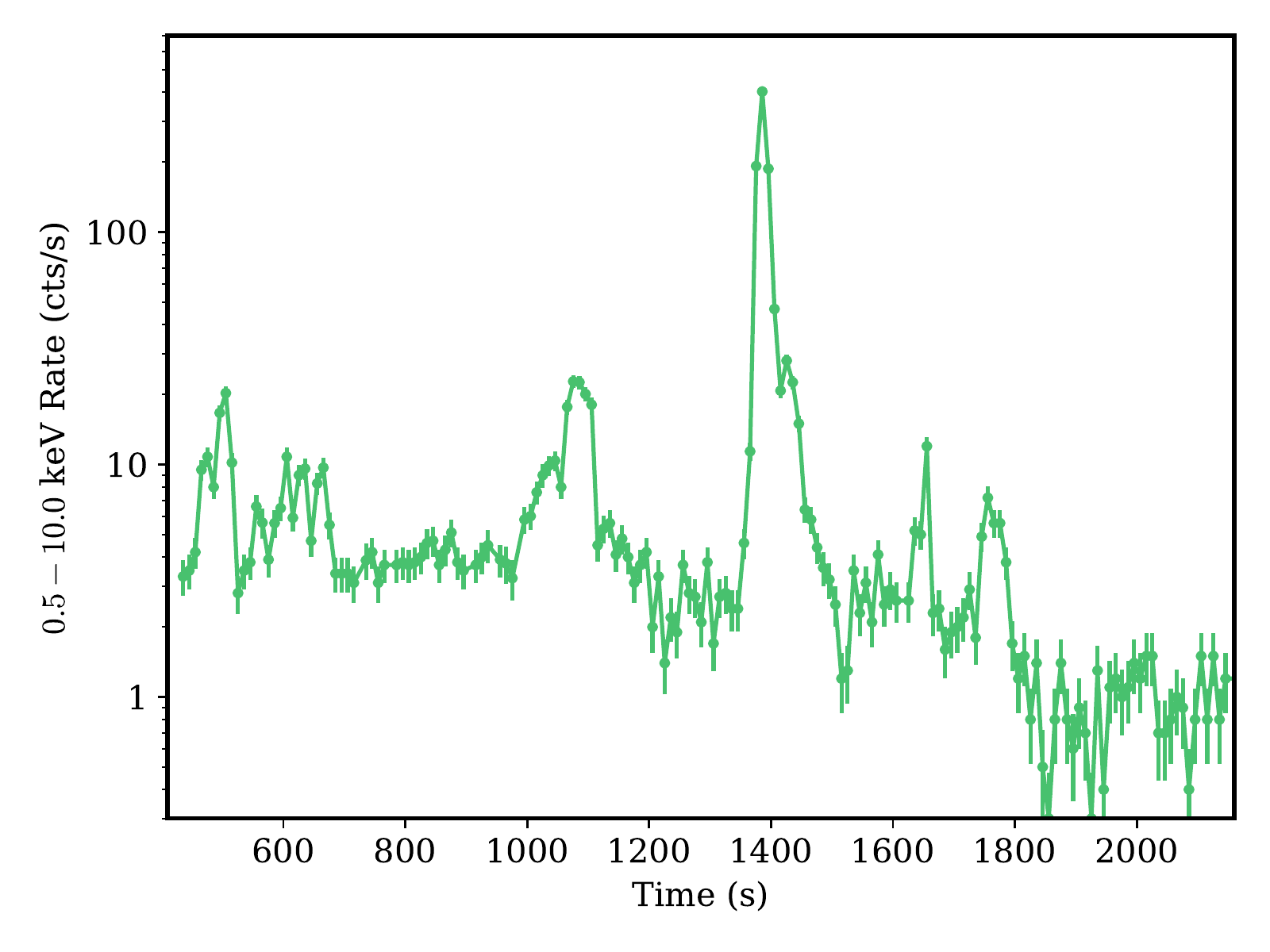}
\includegraphics[width=0.497\textwidth]{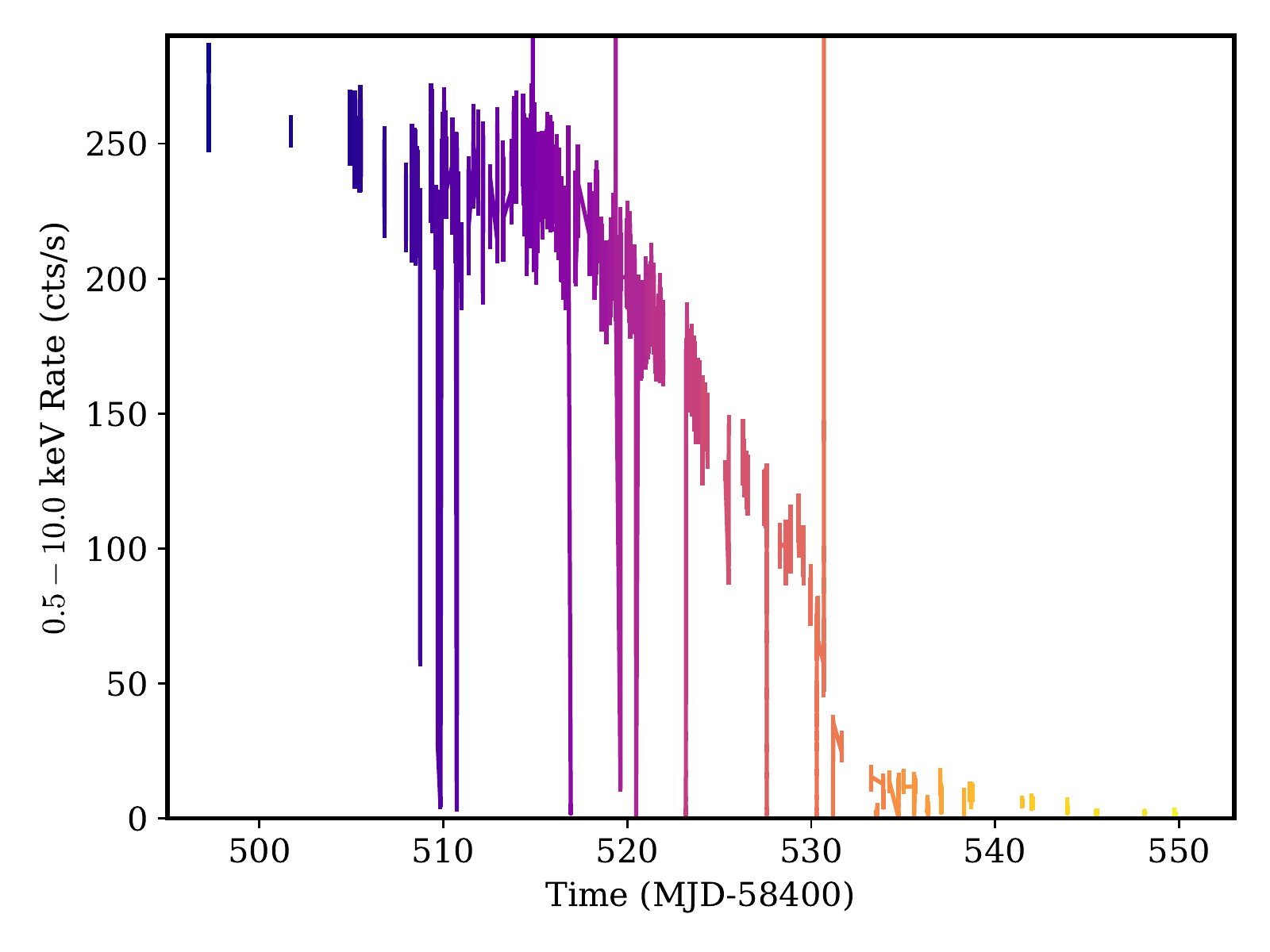}
\caption{{\it Top:} $0.5-10$\,keV \nicer\ light curve of the outburst of \seight, with 10\,s bins (bold colours). Colour indicates time and is matched in the remaining figures. The $4-10$\,keV \maxi\ light curve with 2 day bins is shown in light grey during the \nicer\ data gap due to Sun constraint; we use the change in this flux as the date of the change in state (although \maxi\ does not detect individual flares).
{\it Bottom left:} Zoom to a short section (during MJD 58684) of the 2019 flaring data, showing some strong flares (note the logarithmic scale of the y axis).
{\it Bottom right:} Light curve of recent, steady, state of outburst.
High points during the steady state are due to X-ray bursts; these extend beyond the upper limit of the y-axis and are analysed in detail in \citet{buisson20bursts}.}
\label{fig:lc}
\end{figure*}

\section{Introduction}

X-ray binaries, in which a compact object (a neutron star or black hole) accretes from a secondary star, are important laboratories for studies of accretion and strong gravity. The inclination of the system has an important effect on its observational properties.

There are various ways to measure the binary inclination.
Comparing properties of the two sides of the jet \citep[e.g.][]{hjellming81,mirabel99}, line profiles in X-ray spectra \citep[e.g.][]{fabian89,tanaka95} and the shape and temperature of soft state X-ray emission \citep[e.g.][]{parker19} all give measurements of the inclination of material close to the compact object. Polarimetric X-ray measurements will also be able to determine the inclination of this material \citep{li09}.
Somewhat larger radii are probed by optical and X-ray disc winds, which are most commonly observed at high inclinations \citep[e.g.][]{ponti12,diaztrigo16,higginbottom18}, so observations of these winds can suggest a high inclination of the disc from which they are launched.
The inclination of the binary orbit itself can be constrained by models of the optical brightness variation through the orbit \citep[and references therein]{orosz14}, although this can only be done once the source has reached quiescence.

The most robust indicator of a binary being viewed at high inclination is the presence of periodic eclipses in the light curve due to the X-ray emitting region being occulted by the secondary star \citep[e.g.][]{cominsky84,parmar86,frank87,arzoumanian94}. However, deriving the exact inclination from the eclipse length requires additional knowledge of the size of the companion star and only relatively few systems are at high enough inclination to show eclipses. The minimum inclination to view eclipses depends on the mass ratio and is higher for systems with a lower mass companion. On a similar principle, periodic absorption dips also suggest a high inclination.
Although rare, eclipsing binary systems provide the cornerstone of physical property measurements, as well as routes to understanding the physics of mass transfer, the properties of the donor star atmosphere, and those of the disc itself.

\subsection{\eighteen}

\eighteen\ is a low mass X-ray binary (LMXB), discovered as an X-ray transient in October 2018 \citep{krimm18}, with a variable optical counterpart \citep{vasilopoulos18,baglio18}.

\eighteen\ was notable initially for having unusually strong variability in its X-ray emission \citep[Fogantini et al. in prep.]{ludlam18_1858,hare20}, by factors of several hundred within a few hundred seconds.
This level of variability is rare, having been seen previously in only a few sources \citep{koljonen20}.
It was the dominant observational state for outbursts of \vsgr\ \citep{wijnands00,revnivtsev02} and \vcyg\ \citep{zycki99,walton17,motta17abs}, while GRS~1915+105 has recently shown similar properties \citep{homan19grs,neilsen20}. These other highly variable sources all host black holes (\vsgr: \citealt{orosz01}; \vcyg: \citealt{casares92}, GRS~1915+105: \citealt{greiner01}).

More recently, \eighteen\ showed a more steady flux \citep{buisson20atel1}; during this phase, several Type~I X-ray bursts were detected \citep{buisson20bursts}, identifying \eighteen\ as a neutron star system, although pulsations have not been detected. \seight\ faded during this phase and became undetectable to \nicer\ during the middle of 2020.

The steady flux also allowed for the detection of strong periodic drops in flux, consistent with eclipses by the secondary star \citep{buisson20atel2}, which are the focus of this work.
We present the available observations and their reduction in Section~\ref{sec:odr}; describe our results in Section \ref{sec:res}; and consider their implications in Section~\ref{sec:dis}.

\section{Observations and data reduction}
\label{sec:odr}

We consider all available \nicer\ \citep{gendreau16} data of \eighteen; this is all OBSIDs starting 120040, 220040, 320040 or 359201 and covers times from 2018 Nov 01 to 2020 July 08.

We reduce the data using the standard \nicer\ pipeline, leaving most filtering criteria at their standard values. To include data taken at low Sun angle, where optical loading is comparatively high, we relax the undershoot rate limit to allow up to 400\,cts/s (per FPM). Optical loading degrades the response and elevates the background at low energies ($\lesssim0.4$\,keV) but this is not a problem for this work because we do not perform detailed spectral analysis and we exclude the lowest energies ($<0.5$\,keV) from the analysis.
We also filter out times of high background by requiring a 12-15\,keV (where there is minimal effective area to source photons) rate $<0.3$\,cts/s for all FPMs combined \citep[e.g.][]{bult18}.
We barycentre events to the ICRS reference frame and JPL-DE200 ephemeris and extract spectra and light curves using \textsc{xselect}.
We use the 3C50 model (version~6, Remillard et al. submitted)\footnote{\href{heasarc.gsfc.nasa.gov/docs/nicer/tools/nibackgen3C50\_README.txt}{heasarc.gsfc.nasa.gov/docs/nicer/tools/\\nibackgen3C50\_README.txt}} to estimate the background spectrum and subtract the rate for the relevant energy ranges from the light curves.

We find that the edges of some good time intervals (GTIs) show achromatic dips which occur at different times in different regions of the detector plane. These are due to occultation of the detector plane by parts of the ISS \footnote{\href{https://heasarc.gsfc.nasa.gov/docs/nicer/analysis_threads/iss_obstruction/}{https://heasarc.gsfc.nasa.gov/docs/nicer/analysis\_threads/\\iss\_obstruction/}}; while such times should be filtered out by the \nicer\ pipeline, some differences between the ISS and its model may result in occultations being included in GTIs.
To remove these, we extract light curves for the upper/lower and left/right sides of the detector plane separately (by using \textsc{nicerclean} with the filter expression "RAWX<4" or "RAWY>=4" etc.) and flag times when either of these pairs differ by $>10$\%.

These filters leave a total good exposure time of 345\,ks, of which 210\,ks is in the flaring state (2018-9) and 135\,ks is from the steady state (2020).

We also make qualitative comparisons to the long-term light curve from \maxi\ \citep{matsuoka09}, which we obtain from the online service\footnote{\href{http://maxi.riken.jp/star\_data/J1858-082/J1858-082.html}{http://maxi.riken.jp/star\_data/J1858-082/J1858-082.html}}.

\section{Results}
\label{sec:res}

\begin{figure}
\includegraphics[width=0.5\textwidth]{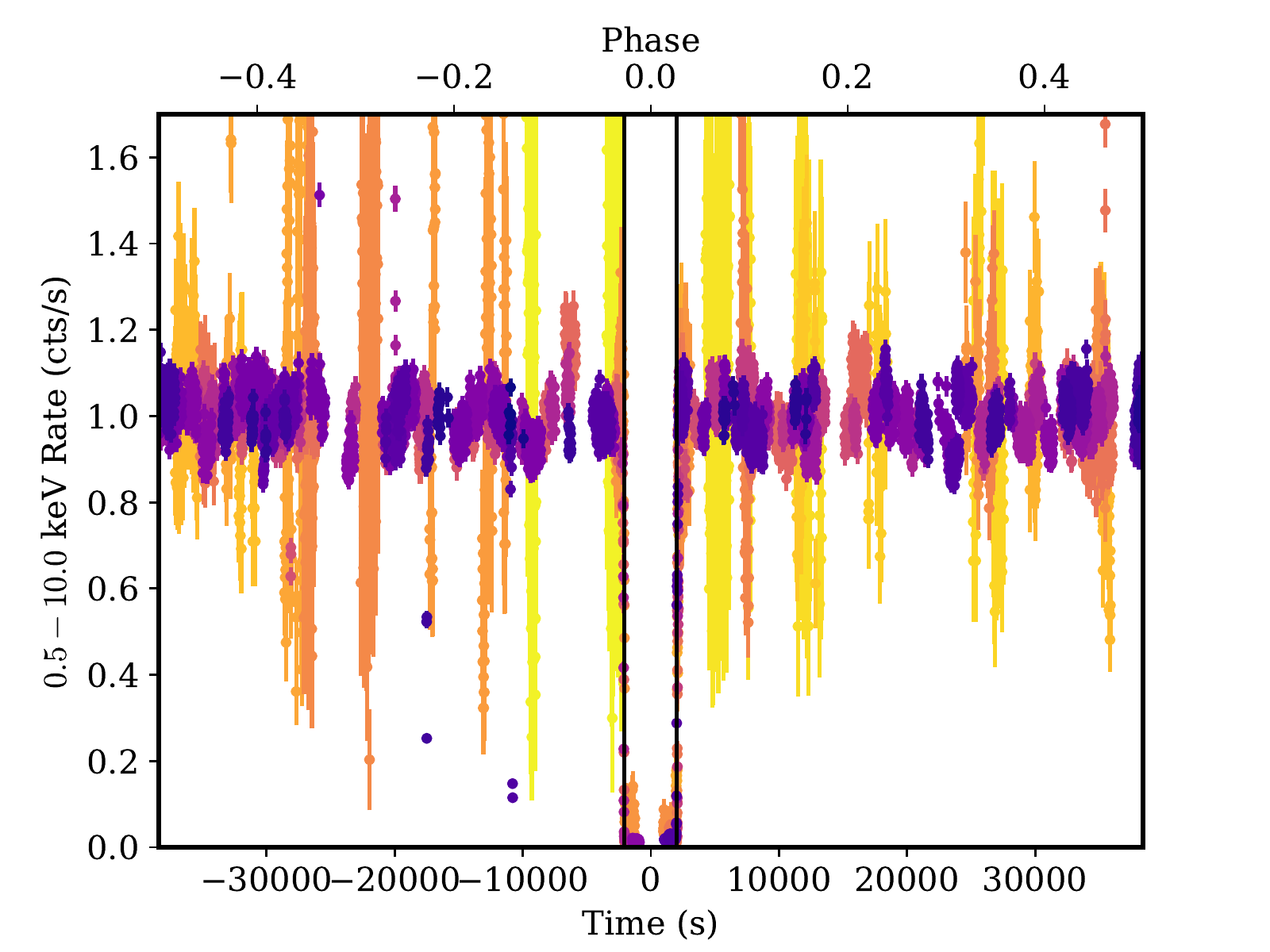}
\caption{X-ray light curve of 2020 data of \seight, as a ratio to a smooth spline and folded on the best fit orbital period (76841.3\,s). The vertical black lines mark the edges of the eclipse (detail in Figure~\ref{fig:lc_eg}), within which the count rate is always low. Some dips are also seen at phases before the eclipse ($-30000$ to $-10000$\,s, detail in Figure~\ref{fig:dips}).
}
\label{fig:fold_lc}
\end{figure}

\begin{figure*}
\includegraphics[width=0.49\textwidth]{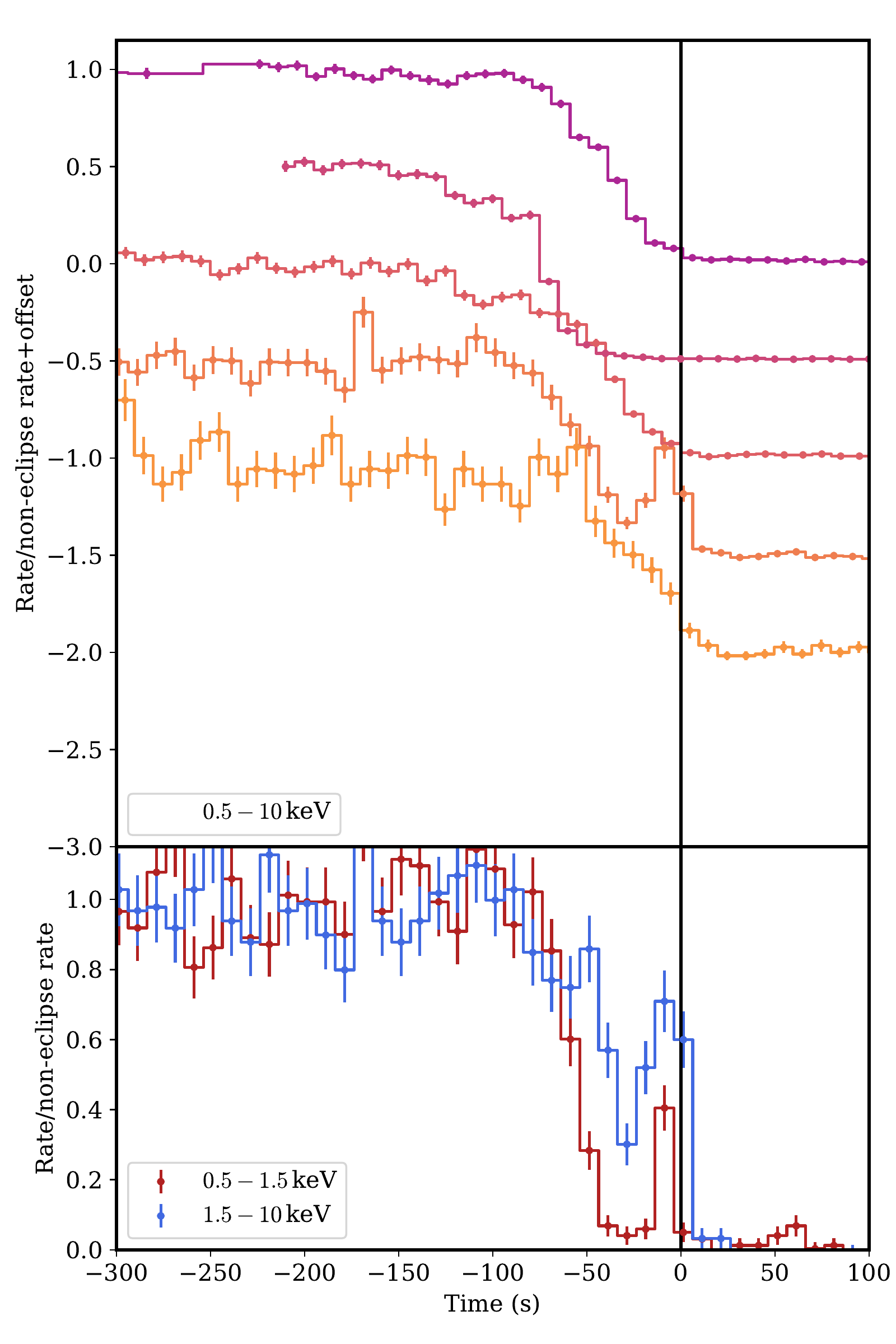}
\includegraphics[width=0.49\textwidth]{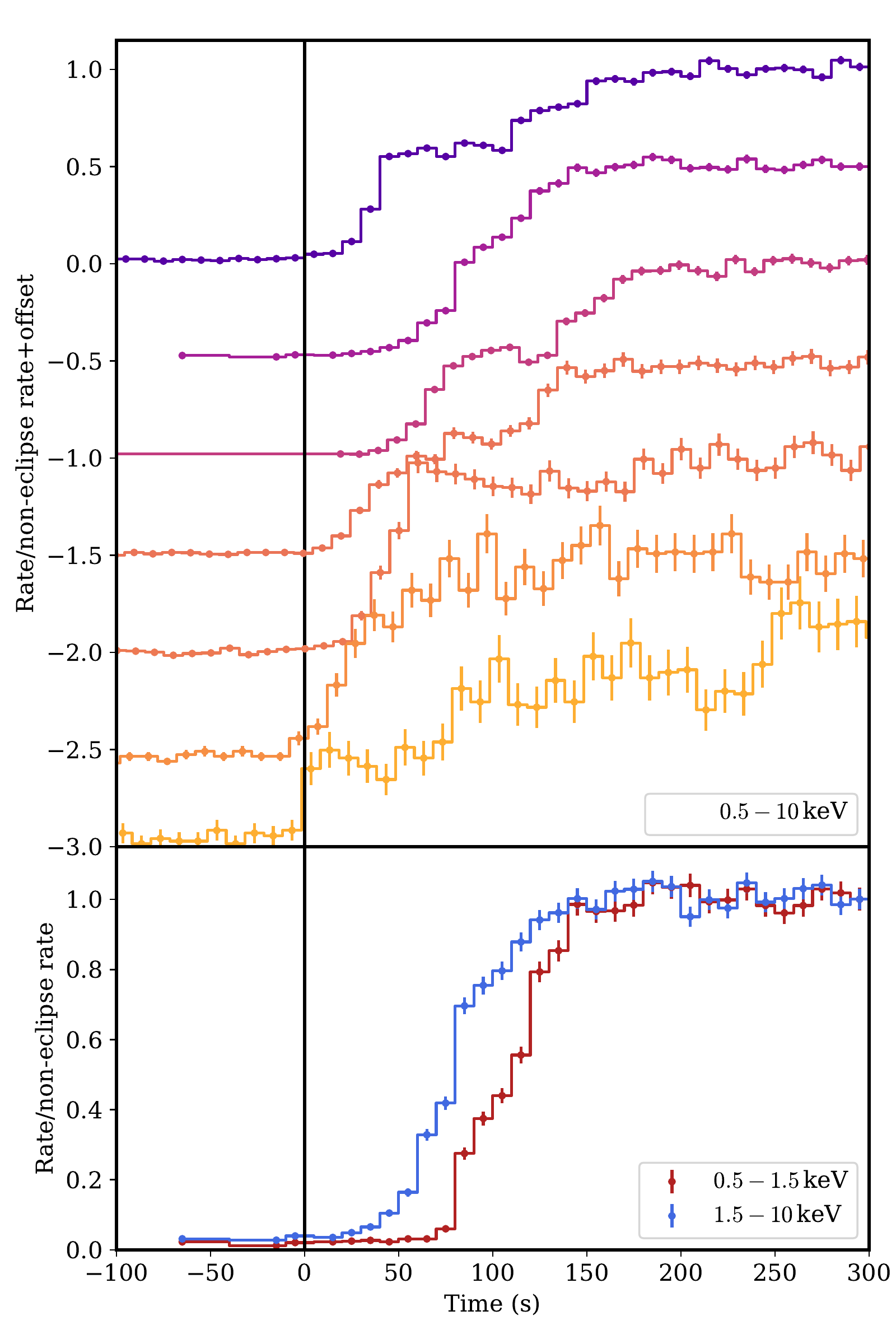}
\caption{X-ray light curve of \seight\ folded on the proposed period, zoomed to eclipse ingresses and egresses. The transitions are extended and show structure which differs between orbits.}
\label{fig:lc_eg}
\end{figure*}

\begin{figure}
\includegraphics[width=0.5\textwidth]{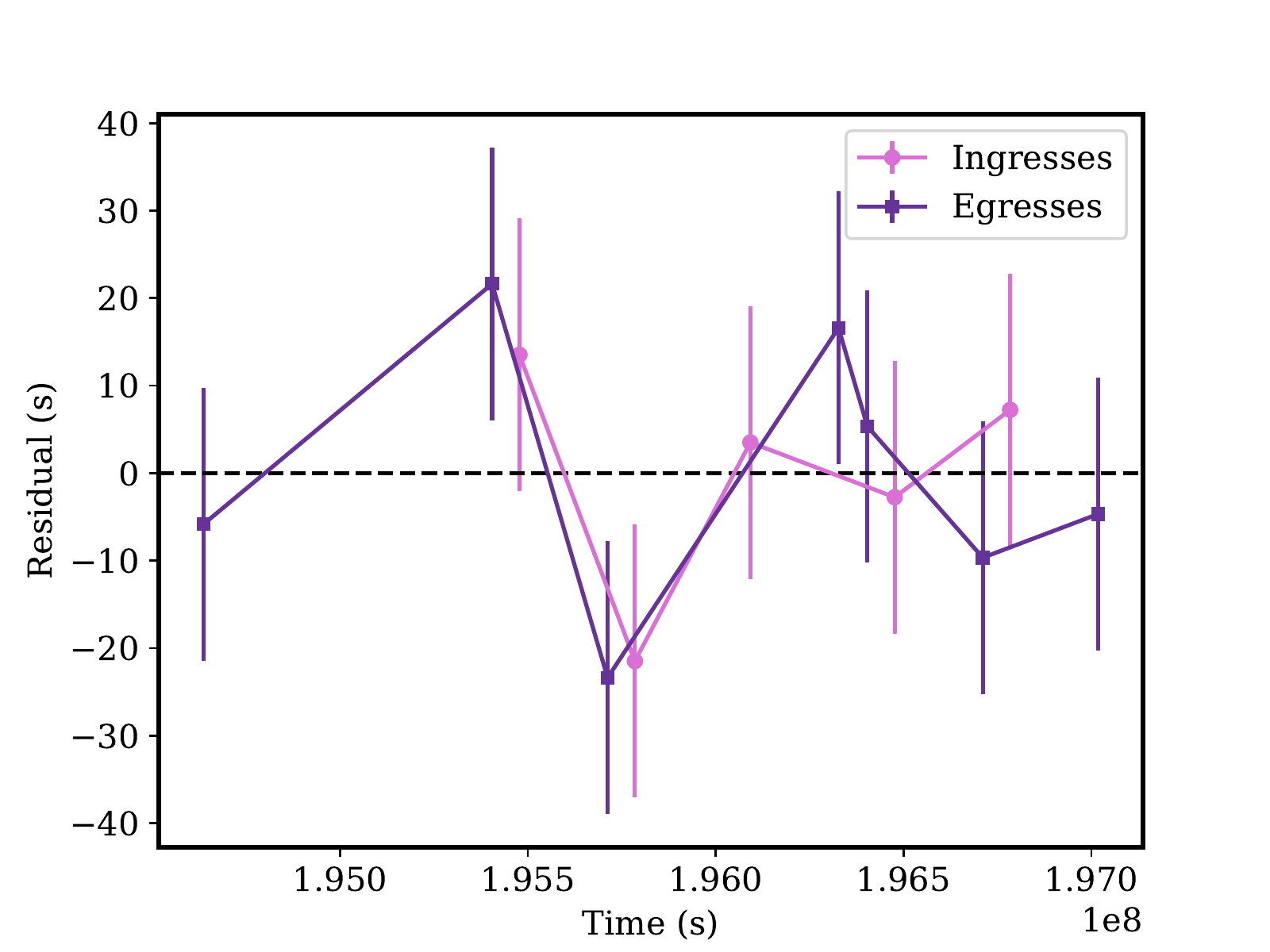}
\caption{Residuals of eclipse times to best fitting period for \seight. Errorbars indicate the scatter in the observed times.}
\label{fig:fit}
\end{figure}

\begin{figure*}
\includegraphics[width=\textwidth]{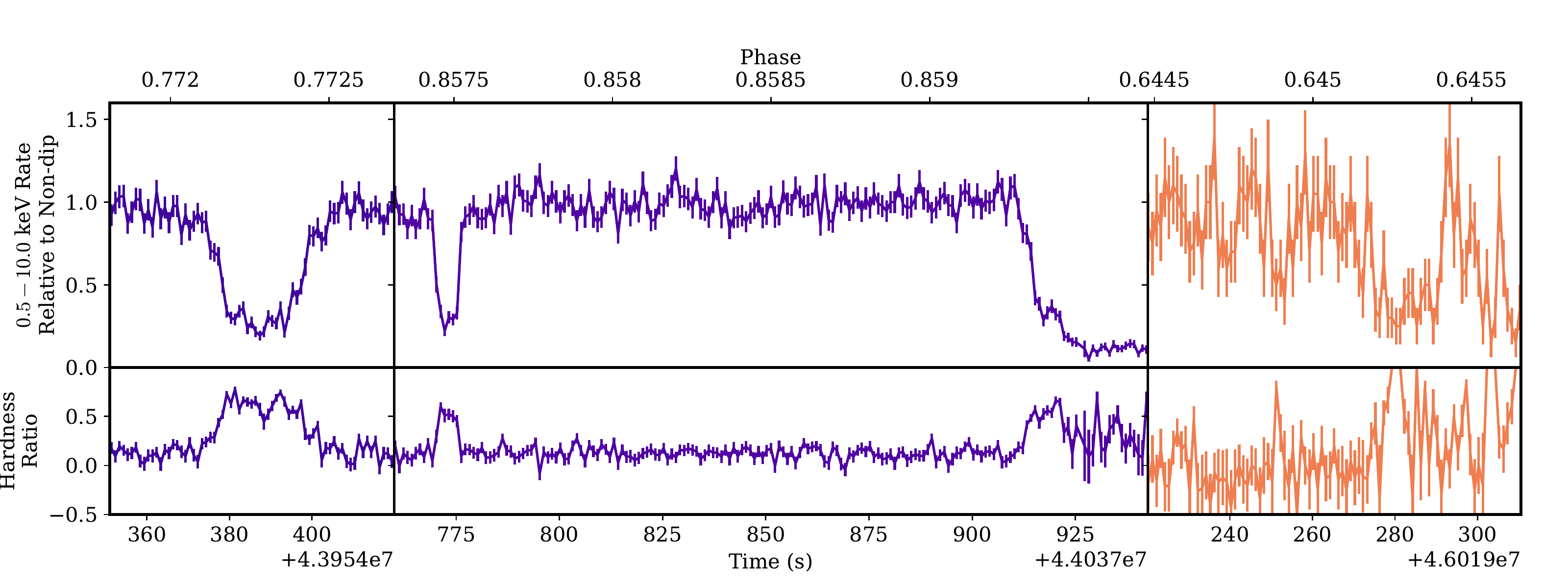}
\caption{Dips in the X-ray light curve of \seight\ which are not due to eclipses. The upper panels show the rate relative to the mean rate outside the dip. The hardness ratio is $(H-S)/(H+S)$ with $H$ the $1.5-10$\,keV and $S$ the $0.5-1.5$\,keV rate. These are associated with an increase in hardness, as would be expected from absorption \citep{diaztrigo06}.}
\label{fig:dips}
\end{figure*}

The light curves of \eighteen\ at various times and time scales are shown in Figure~\ref{fig:lc}. The latter stage of the outburst (after MJD 58885) shows the highest steady count rate, although there are a number of peaks at a higher rate earlier in the outburst.
The assumption of the high, steady count rate occurred while \nicer\ could not observe due to the proximity of the Sun, but a jump in flux is also seen in the \maxi\ light curve around MJD 58885, so we choose this date to divide the stages of the outburst.
The peaks early in the outburst are irregularly shaped but all show a rapid increase in rate from the typical local value; we refer to them as flares (without implying that they are necessarily due to an increase in primary emission) and this interval (before MJD 58885) as the flaring state.
In the steady state (after MJD 58885), there is a long-term decline but also shorter timescale features.
These include Type~I X-ray bursts, which are described in detail in \citet{buisson20bursts}, and various dips.
Some of the dips are very brief, while others are deep and prolonged. Additionally, some observation sections consist entirely of a much lower count rate than adjacent sections, similar to nearby dips, appearing to consist only of time during a dip.
The deep dips occur at regular intervals -- aligning adjacent dips by folding the light curve on an appropriate period (Figure~\ref{fig:fold_lc}, the precise value will be found in Section~\ref{sec:per}) shows that all dip in/egresses line up well and the in-dip phase always shows a low count rate.
For visualisation purposes, we show the folded light curve as a ratio to a smooth B-spline \citep{dierckz75} fitted away from the eclipses and with smoothing parameter chosen to fit the long-term shape of the light curve.
This period is robust to gross changes (e.g. half, double etc.):
there are observations of adjacent eclipses, so the period cannot be longer; and
the 1/2, 1/3, etc. phases are shown to be bright in the folded light curve (Figure~\ref{fig:fold_lc}), so it cannot be shorter.
The most likely explanation for this is that the deep dips are due to eclipses by the secondary star in the binary. We proceed to calculate quantitative parameters for this situation.

\subsection{Ingress and egress structure}
The data include 5 ingresses and 7 egresses; light curves of the in/egresses are shown in Figure~\ref{fig:lc_eg}.
From this, it is apparent that the in/egresses are extended in time, lasting around 100\,s. Egresses may be slightly  longer, taking up to 200\,s to achieve the full non-eclipse flux.
Also, the eclipses last longer at lower energies. This can occur with a progressively increasing absorbing column density, since soft flux is obscured by a lower column density, which occurs further from the centre of the star.
Furthermore, the in/egresses are each slightly different in shape, including one example where the flux drops and recovers before the true ingress.
These differences may be due to variable structure in the surface layers of the star, although detailed consideration of the physics of the stellar atmosphere is beyond the scope of this paper.

\subsection{Period}
\label{sec:per}

We calculate the period, and simultaneously the eclipse duration, by fitting to the observed times of ingress and egress.

We define the start of an eclipse as the end of the flux drop during ingress: the earliest point in the light curve (binned to 10\,s) consistent at 1-$\sigma$ with the mean rate after this point; the end of an eclipse is defined equivalently after time-reversal.
We then use a least-squares fit to these points of a model with constant period, eclipse duration and eclipse phase, fitting for the period, eclipse duration and ingress time (or equivalently period, ingress time and egress time).
We use the empirical dispersion of the measured in/egress times to estimate their error (equivalent to setting the reduced $\chi^2$ to 1): we give each measurement the same error, $\sigma$, choosing $\sigma$ such that
$$\sum_{i=1}^{N}{\frac{\left(t_{{\rm meas}, i}-t_{{\rm pred}, i}\right)^2}{\sigma^2\left(N-M\right)}}=1$$
where there are $N$ measured times (ingresses or egresses), $t_{{\rm meas}, i}$, each with prediction from the model $t_{{\rm pred}, i}$ and $M=3$ is the number of free parameters. This gives $\sigma=16.5$\,s; we note that this is significantly more than the dispersion expected from the light curve binning (which is $\approx2.9$\,s).
Residuals to this model are shown in Figure~\ref{fig:fit}.
We then find confidence limits on the best-fit values in the standard way, solving for a given change, $\Delta\chi^2$, in
$\sum_{i=1}^{N}{\frac{\left(t_{{\rm meas}, i}-t_{{\rm pred}, i}\right)^2}{\sigma^2}}$
when varying a given parameter; for a 90\% confidence limit, $\Delta\chi^2=2.71$ \citep{lampton76}.
The fitted values and confidence intervals are period
$P=76841.3\pm1.4$\,s and eclipse duration
$t_{\rm ec}=4098\pm18$\,s.
Mid-eclipse of the best constrained orbit (at the middle of the observations) occurs at \nicer\ mission time (TDB) $195786918.1\pm0.5$\,s, MJD 58924.052269; hereafter, we define phase such that mid-eclipse occurs at phase 0.

\subsection{Non-eclipse dips}
\label{sec:dips}

We also find several dips which do not align with the eclipse phase (Figure~\ref{fig:dips}). These dips are not as deep as the eclipses: the in-dip rate is at least 10\% and typically around 25\% of the adjacent non-dip rate, compared to at most a few per cent for the eclipses. These dips show an increase in hardness, so are likely due to absorption by a thickened region in the accretion flow; this is typical of dipping LMXBs \citep{diaztrigo06}. Additionally, the deepest parts of some dips show a re-softening, as has been seen in some other sources \citep{kuulkers98,tomsick98}. This may be due to the absorption becoming strong enough to absorb hard as well as soft X-rays from the primary source, leaving only softer scattered emission. All of the observed dips occur at phases before the eclipse, from 0.64 to 0.86.
Detailed consideration of the spectral changes during the dips is beyond the scope of this work; the salient point here is that all dips occur during the pre-eclipse phase of the orbit, as expected if they are due to the raised rim of the disc in the pre-eclipse intervals.

\subsection{Eclipses in the flaring state?}

\begin{figure}
\includegraphics[width=0.5\textwidth]{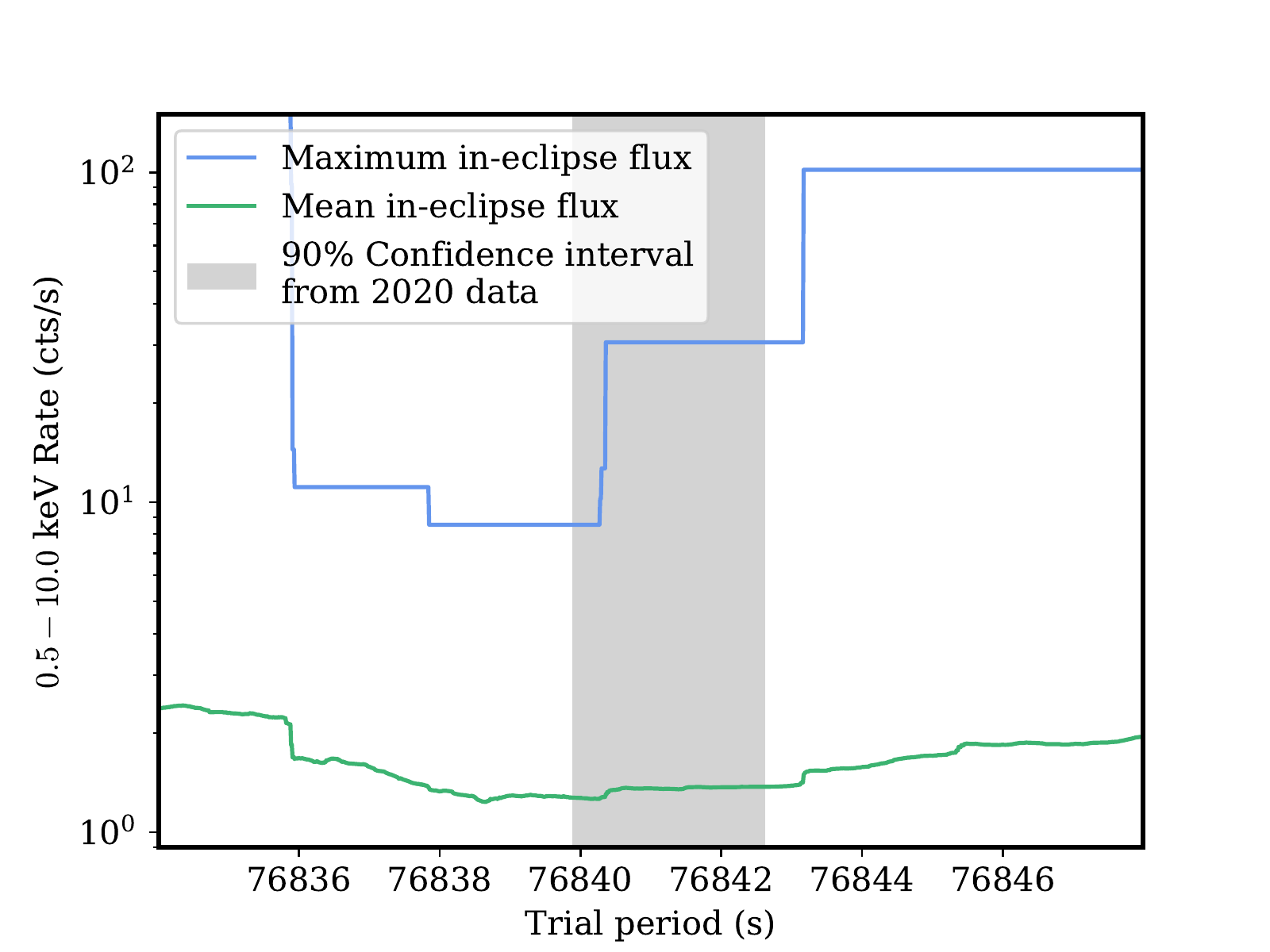}
\caption{Mean and maximum flux during the eclipse in the flaring state data of \seight, as a function of assumed period. The mean in-eclipse flux is low and there are no bright flares during the eclipse when using a period close to the best fit from the steady state data.}
\label{fig:flaring_eclipse}
\end{figure}

Unless the binary orbit was significantly perturbed, the period and eclipses seen in the 2020 data should also be present in the earlier flaring data, although they were not noticed at the time.
To test this, we calculate the flux observed during eclipses, computing the eclipse times from ephemerides allowed by the 2020 data.
We fold the light curve of 2018/9 data on each of a range of periods around that found in the 2020 data and for each period calculate the mean and maximum count rate during the eclipse (defined by the extrapolation of the best fit ingress and egress times).
We show the resulting count rates as a function of trial period in Figure~\ref{fig:flaring_eclipse}. The in-eclipse flux is low when folding on periods supported by the 2020 data: the maximum flux during the eclipse when folding on a trial period within 76838-76840.5\,s is 9.5\,counts/s; and the mean flux for periods of 76838-76842\,s is less than 1.3\,counts/s.
The mean flux for the full 2018-9 dataset is significantly higher, 3.7\,cts/s.

This shows that, as would be expected, no large flares occur when the system was in eclipse. However, the in-eclipse rate does reach a peak of 9.5\,cts/s, implying some variability.
Therefore, we compare the fractional variability of the observed rate during in-eclipse and non-eclipse phases.
The fractional variability is defined as $F_{\rm Var}=\frac{\sqrt{{\rm Var}(x)-\langle\sigma^2\rangle}}{\langle x\rangle}$ \citep[e.g.][]{nandra97,vaughan03}, where $x$ are the observed data values, $\sigma$ their errors, ${\rm Var(\cdot)}$ is the variance and $\langle\cdot\rangle$ the mean.
We use the $0.5-10$\,keV energy range and 10\,s bins. We find that the variability is much higher during non-eclipse phase, with $F_{\rm Var}=3.50\pm0.02$ compared to $F_{\rm Var}=0.498\pm0.008$ during eclipses. This latter value is much more typical of XRBs in the hard state \citep[e.g.][]{belloni05}.
This can be explained by scattering of the emission by the optically thin atmosphere and wind of the star or by a large scale accretion disc corona \citep[e.g.][]{parmar86} or disc wind material.

This is also not an unreasonably high scattered brightness.
By comparing the in-eclipse with out-of-eclipse spectra from 2020, we find that the in-eclipse rate is around 2\% of the out-of-eclipse rate. Applying the same factor to the peak in-eclipse rate gives a peak rate of $\sim500$\,cts/s, which is not unprecedented in the light curve of \eighteen: several flares reach 600-700\,cts/s and Type~I X-ray bursts reach $\sim1500$\,cts/s \citep{buisson20bursts}.

We also place an upper limit on the mean period change during the outburst $-1.5\times10^{-7}<\dot{P}<3\times10^{-7}$, by testing a linear change in period and requiring that no flares over 25\,cts/s occur during a predicted eclipse. Compared to measurements of period changes in NS XRBs, of order $10^{-12}$ \citep[e.g.][]{patruno12period}, this is not a strong constraint.

\subsection{Phase dependence of flares}

\begin{figure}
\includegraphics[width=0.5\textwidth]{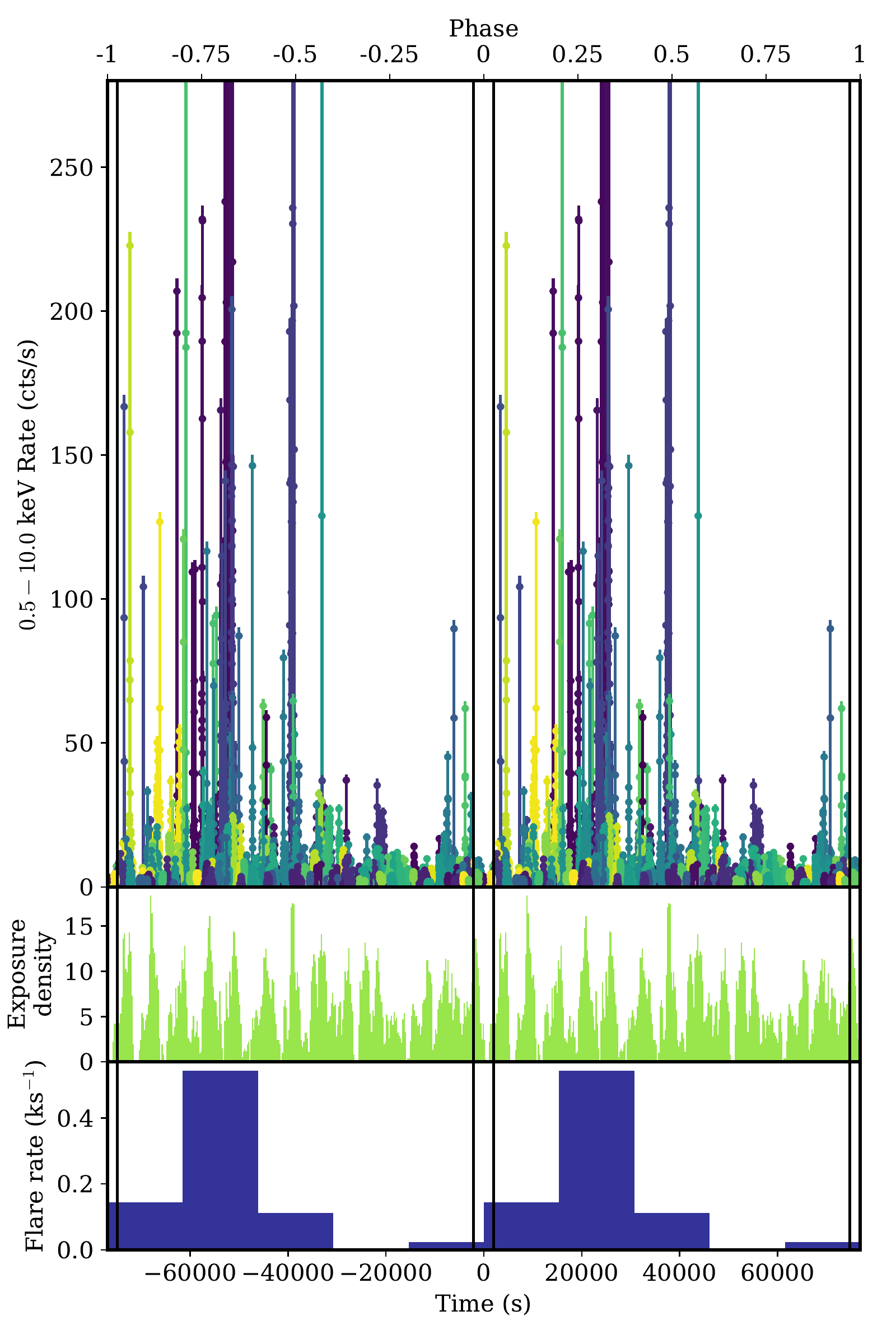}
\caption{\textit{Top:} X-ray light curve of flaring state (2018-9) data (colours) folded on the orbital period from the 2020 data. No bright flares are seen during the eclipse phase and flares are stronger and more common in the post-eclipse than pre-eclipse half of the orbit.
\textit{Middle:} The distribution of exposure time across orbital phase (number of exposures at a given phase) shows fine structure but is broadly uniform across phase.
\textit{Bottom:} The number of flares per exposure time is strongly non-uniform, peaking after the eclipse and having very few flares before the eclipse.
}
\label{fig:fold_lc_19}
\end{figure}

Folding the 2018-9 flaring state data also appears to show more flaring activity at phases after the eclipse than before (Figure~\ref{fig:fold_lc_19}, lower panel). This is not due to different amounts of exposure at different phases, as the exposure is distributed quite evenly throughout the orbit (Figure~\ref{fig:fold_lc_19}, middle panel).
To test whether this inhomogeneity is real or could be due to chance, we compare the observed flare distribution with that expected if they were random, by calculating the flare rate in 5 equal phase bins (Figure~\ref{fig:fold_lc_19}, upper panel).
We define a flare as having a peak rate of $\geq80$\,cts\,s$^{-1}$ and requiring that the rate drop below 10\,cts\,s$^{-1}$ between  different peaks for them to count as separate flares.
We find that most flares (20 of 28) occur between phases of 0.2 and 0.4 (where the eclipse midpoint is 0);
there is a $3.6\times10^{-10}$ chance of this bin having at least this many flares if the flares were distributed uniformly in exposure.
Therefore, we conclude that more flares occur at phases following the eclipse.

\subsection{System and Companion Star Parameters}

\begin{figure}
\includegraphics[width=0.49\textwidth]{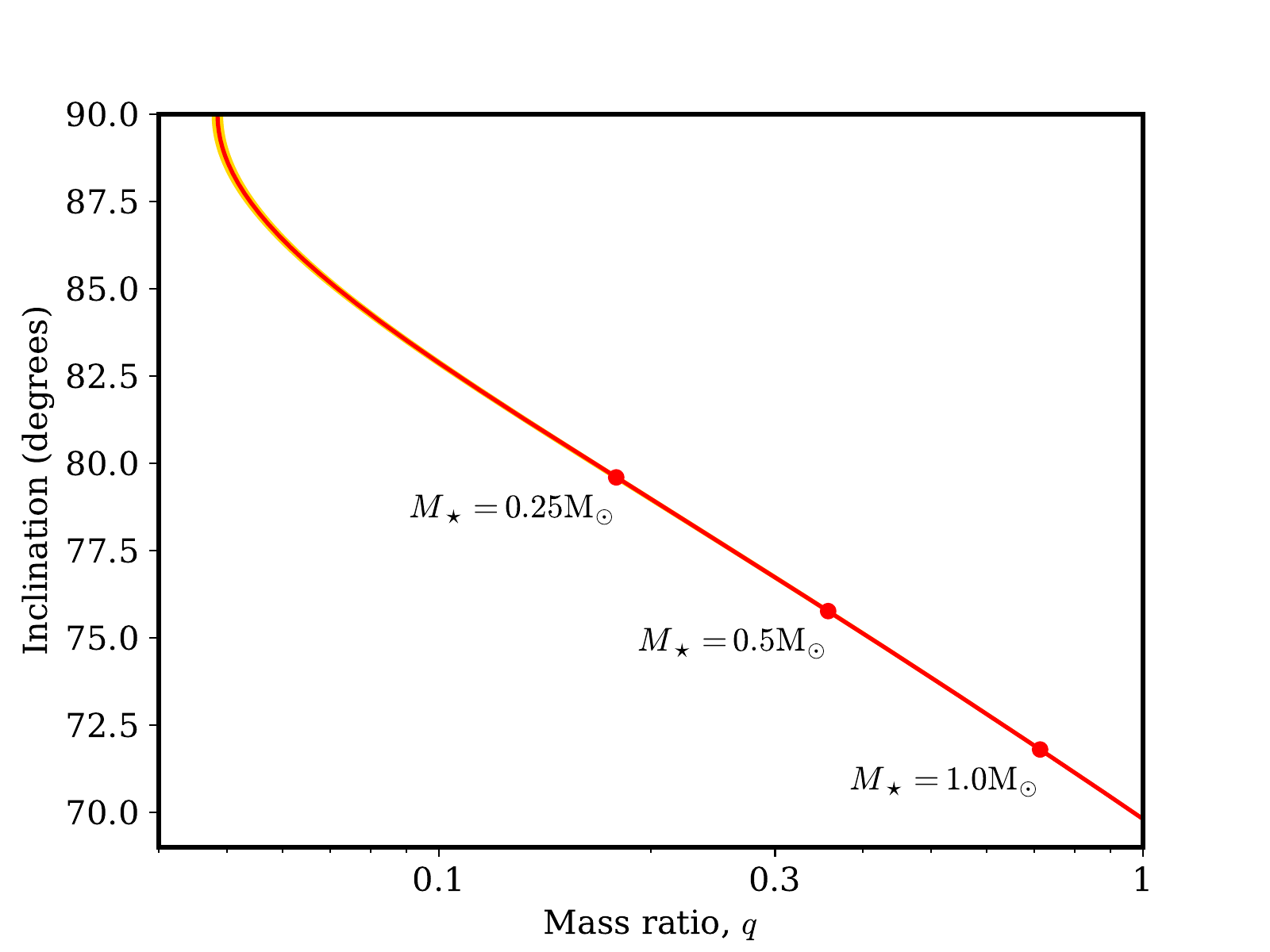}
\caption{Relation between inclination and mass ratio for \seight. The values for the best fit period and eclipse duration are shown in red; the statistical uncertainty gives a $1\sigma$ confidence interval (gold) similar to the width of the line. Example companion stellar masses ($M_{\rm \star}$) are given for $M_{\rm NS}=1.4\,{\rm M_{\odot}}$; at a given point in the plane, $M_{\rm \star}$ is proportional to the assumed $M_{\rm NS}$.}
\label{fig:incl}
\end{figure}

\begin{figure}
\includegraphics[width=0.49\textwidth]{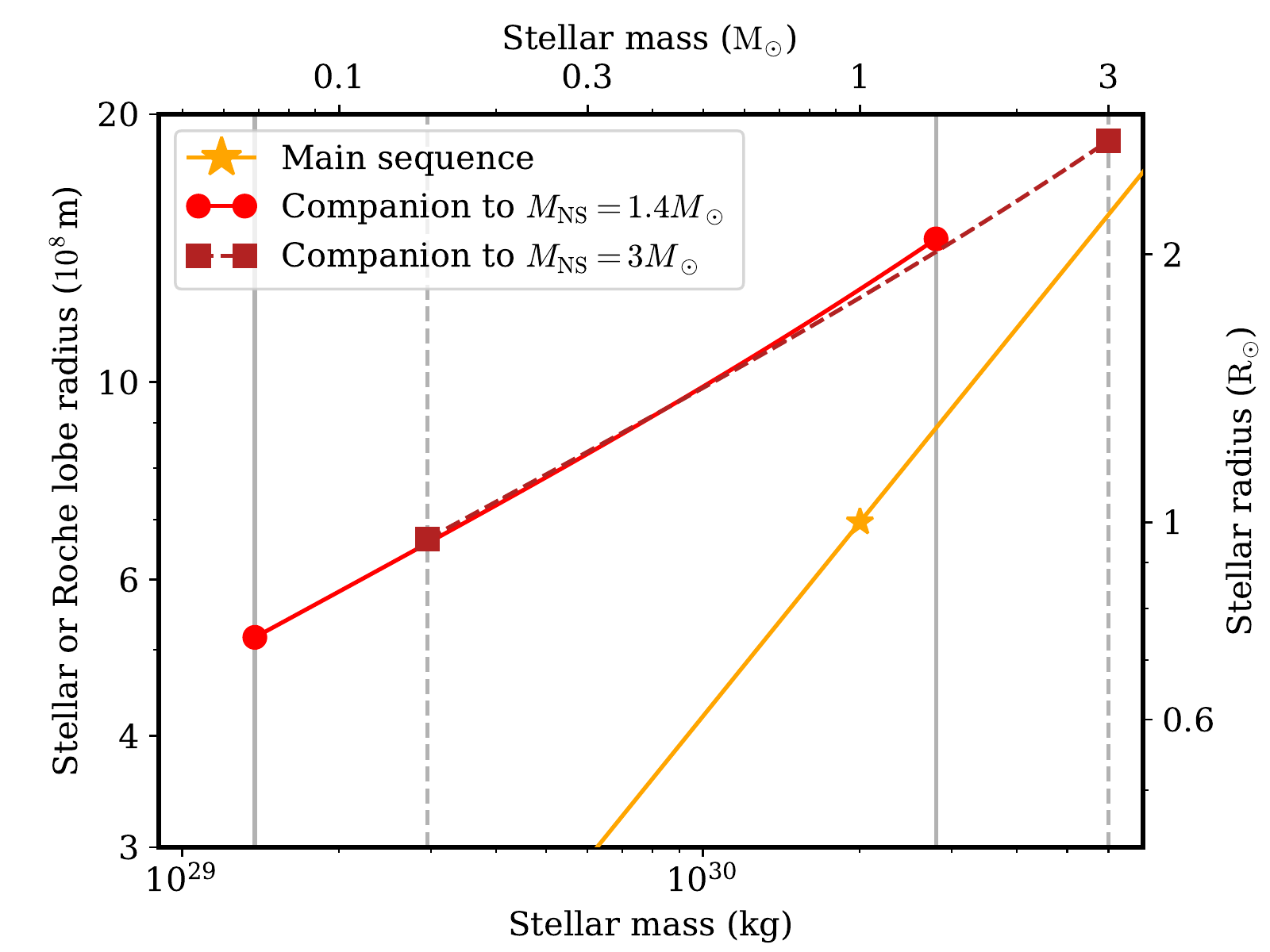}
\caption{Mass-radius relation for possible companion stars in \seight\ implied by the orbital period (red). This has only a very weak dependence on the neutron star mass (shown for: $M_{\rm NS}=1.4\,{\rm M_{\odot}}$, lighter, solid, circles; and $M_{\rm NS}=3\,{\rm M_{\odot}}$, darker, dashed, squares). The marked line ends indicate the mass constraints from requiring $M_{\star}\leq M_{\rm NS}$ and the minimum implied by the eclipse length. We also show the mass-radius relation for main sequence stars (yellow, with Sun marked), clearly demonstrating that the donor in \seight\ is evolved (a likely sub-giant).}
\label{fig:mr}
\end{figure}

The presence of eclipses requires a high inclination; the exact value depends on the radius of the secondary star relative to the orbital separation, which is governed by the mass ratio. Assuming a circular orbit gives
$$r_{\rm L}(q) = \sqrt{\cos^2 i + \sin^2 i  \sin^2 \left(\frac{\pi t_{\rm ec}}{P}\right)}$$
where $r_{\rm L}(q)$ is the Roche lobe radius as a fraction of the orbital separation, which is a function of the mass ratio $q$ \citep{joss84}. The function $r_{\rm L}(q)$ is not soluble analytically but has good approximations \citep[e.g.][]{paczynski71, eggleton83}; we use that of \citet{eggleton83}.
The resulting relation between mass ratio and inclination is shown in Figure~\ref{fig:incl}.
Various physical considerations limit the possible range of inclination. Requiring that the companion star  have lower mass than the neutron star gives $q<1$ and $i\gtrsim70^\circ$. If the disc is aligned with the binary orbit, the finite thickness of the disc implies an inclination of less than $90^\circ$; the exact limit depends on the thickness of the disc and the geometry of the emitting region. For example, a typical $H/R=0.1$ and a small emitting region gives $i<84^\circ$.

The orbital period also determines the companion's mass as a function of radius; this is shown in Figure~\ref{fig:mr}. Using the main sequence relation of \citet{demircan91}, we find that the donor star is larger than a main sequence star of the same mass by a factor of a few (dependent on the exact mass). This difference is present even with a conservative upper limit for the neutron star mass, $M_{\rm NS}=3\,{\rm M_{\odot}}$, which is higher than any empirically confirmed mass \citep{antoniadis13} and at the upper end of theoretical predictions \citep{lattimer01}.
Therefore, the donor in \seight\ is a sub-giant.

\section{Discussion}
\label{sec:dis}

We have discovered eclipses in the \nicer\ monitoring of \eighteen. This shows:

\begin{itemize}
\item The orbital period is $P=76841.3\pm1.4$\,s ($21.345\pm0.0004$\,h).

\item The binary inclination is high, $>70^\circ$.

\item The companion star is larger than a main sequence star of the same mass.

\item The eclipse ingresses and egresses are extended, which appears to be due to absorption in the atmosphere of the stellar companion.

\item Several other absorption dips are also present, all during the pre-eclipse half of the orbit.

\item Flares seen in the earlier, highly variable, observations occur only during post-eclipse phases of the orbit.

\end{itemize}

\subsection{Period}

Being based simply on eclipse ingress and egress times, there are few sources of systematic error in the period measurement, although these times may not occur at precisely the same phase in each orbit.
This is exemplified by the large scatter ($16.5$\,s) relative to the statistical error with which an individual in/egress time can be measured. This scatter is seen in other eclipsing X-ray binaries \citep[e.g.][]{hertz97,wolff02,wolff09} and is most likely due to structure in the stellar atmosphere at column densities where the eclipse starts/ends, which is evident in the differing shapes of the in/egress profiles (Figure~\ref{fig:lc_eg}, see also \citealt{wolff07}).
The scatter in the in/egress times does not appear to follow a long-term trend so is unlikely to introduce a significant bias into our results; the uncertainty the scatter introduces is included in our stated period uncertainty by using the amount of scatter rather than the statistical uncertainty of the in/egress times when calculating the uncertainty of the period.

\subsection{Binary inclination}

Using the eclipses to estimate the inclination as a function of $q$ assumes that the measured eclipse times (when the star has sufficient optical depth that any transmitted flux is insignificant compared to the scattered and background) occur at the Roche lobe radius, which is unlikely to be precise to better than a few per cent.
Additionally, the formula used for the size of the Roche lobe radius is correct in volume only to within a few per cent and the Roche lobe is not perfectly spherical, so the projected line of sight radius may not match the radius calculated from a sphere of the appropriate volume.
The Roche lobe is slightly extended along the axis towards the companion, so the size perpendicular to this will be smaller than the equivalent sphere; hence, the inclination will be slightly under-estimated.
Therefore, the uncertainty on the $q-i$ relation is at least a few per cent, dominated by systematic rather than statistical errors.
However, due to the large range allowed in this relation, ameliorating these systematic uncertainties with a more detailed treatment of the Roche lobe size is not justified until a more precise constraint on either the mass ratio or inclination is obtained.

The detection of eclipses requires a high orbital inclination ($>70^\circ$). This fits with the detection of winds in the optical \citep{munoz20}, which are most often detected at high inclination.
However, reflection spectroscopy during the flaring state found a low inclination \citep[$i<29^\circ$,][]{hare20}.
Similar mismatches have been found for other sources (e.g. compare \citealt{torres19} with \citealt{buisson19} and \citealt{fabian20}; also \citealt{connors19}).
This could either be because of a pronounced inner disc warp, so the inner disc is at a different inclination to the binary orbit, or an under-estimate in the reflection measurement. The difference in inclination ($>40^\circ$) would be an unusually strong warp. The latter case could occur if distant scattering material, possibly including a conical disc wind, is present in the spectrum but included during fitting as part of the relativistic, inner disc, reflector.
Additionally, \seight\ has a complex spectrum with rapidly variable absorption, which will distort the iron line and so make determining inner disc properties from the iron line shape less reliable.

\subsection{Extended in/egresses}

The eclipse in/egresses are also extended in time, lasting 50-100\,s (Figure~\ref{fig:lc_eg}), and show significant structure, including a dip followed by a re-brightening.
We note that, as might be expected, this duration is similar in fractional terms to that observed in \exo\ (another eclipsing/bursting transient X-ray binary), where the period, eclipse and transition are around a factor of 10 faster \citep{parmar86}.
The most likely source for the differing profiles is changing absorption in the upper layers of the stellar atmosphere.
The alternative, that it is due to gradual occultation of an extended emitting region, is disfavoured for several reasons. Firstly, the expected size of the emitter is too small for the in/egresses to last so long. The largest component emitting significant flux is a disc with peak temperature $\approx2$\,keV close to the inner radius. The limit of detection is for material at around 0.1\,keV, which will occur  at $\approx50R_{\rm in}$ (since $r\propto T^{-\frac{4}{3}}$), which at an orbital speed of a few 100\,km\,s$^{-1}$ would last at most a few seconds. A scattering region could be larger, but would be unlikely to carry so much of the flux.
Further, the spectral hardness changes during the eclipses appear to be due to absorption rather than partial occulation: the soft flux drops away first, followed by the hard flux, as expected for an increasing absorption column density. For progressive occultation of a disc (with a larger soft than hard X-ray emitting region), the soft flux should partially drop, followed by the hard, followed by the remaining soft as the following side is occulted. The reverse happens during egress.

Therefore, the in/egress duration constrains the atmosphere of the companion star. For \seight, 50-100\,s corresponds to a projected length of $\sim10^4$\,km over which \nicer\ is sensitive to changes in absorption column density (from around the line of sight value, $3\times10^{21}$\,cm$^{-2}$ to $\sim10^{24}$\,cm$^{-2}$). If the eclipse is close to the limb of the star, this may correspond to a shallower physical depth.
Additionally, the in/egresses are not smooth changes but show structure, in one case dropping by over 80\% and returning close to the pre-eclipse level while entering the eclipse (Figure~\ref{fig:lc_eg}). A similar profile was seen in \exo, persisting for 7 orbits \citep{wolff07}. This was attributed to a stellar prominence (denser material lifted from the surface by magnetic loops) and a similar structure is the most likely candidate in \seight.
The remaining structure could be due to smaller, non-detached prominences or lines of sight which do not pass through a central low density region.
Absorption from material in the outer disc is also possible but, given the variety of in/egress structure and comparative lack of other dips at phases close to the ingress, disc material is unlikely to cause all of the differences.

\subsection{In-eclipse flux}

During the eclipses in \eighteen, some flux is still present, at around 2\% of the non-eclipse flux. This is not unusual among eclipsing X-ray binaries; for example EXO~0748$-$676 has an eclipse flux of around 4\% in the 2-6\,keV band \citep{parmar86}.
Since the primary X-ray source is much smaller than the companion star, it is unlikely that this flux is due to a true partial eclipse of the inner region, with the source only partially occulted by the star. Instead, contributions could come from X-rays reprocessed in a larger scattering medium or directly emitted by the stellar corona.
However, stellar coronae do not have sufficient power to produce the observed X-ray flux \citep{rosner85,bildsten00}.
The scattering material could potentially be an extended disc atmosphere or wind, or the optically thin surface layers of the companion star. However, the stellar atmosphere is too small to scatter sufficient flux.
An extended disc atmosphere has been inferred from its line emission both during eclipses for EXO~0748$-$676 \citep{psaradaki18} and during the flaring state of \eighteen\ \citep{buisson20rgs}.

\begin{figure}
\includegraphics[width=0.49\textwidth]{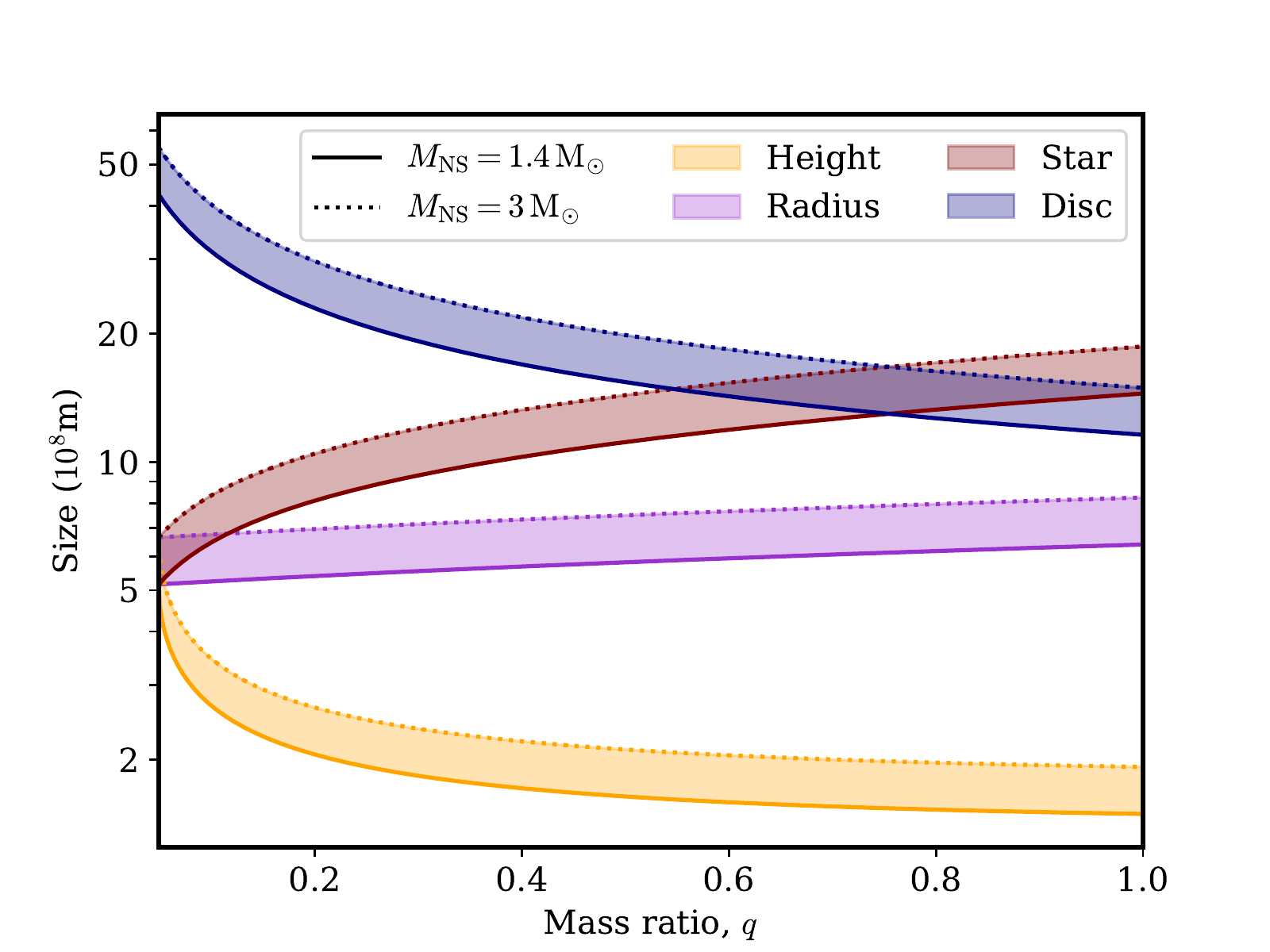}
\caption{Minimum size of the scattering material (which produces the in-eclipse flux) in \seight\ as a function of mass ratio; shaded bands indicate a range of neutron star masses, $1.4-3\rm M_{\odot}$. At least one of the two dimensions  (radius, parallel to the disc plane; or height, perpendicular to the plane) must exceed the given size. For context, the disc and stellar radius are also shown.}
\label{fig:scat}
\end{figure}

\begin{figure*}
\includegraphics[width=\textwidth]{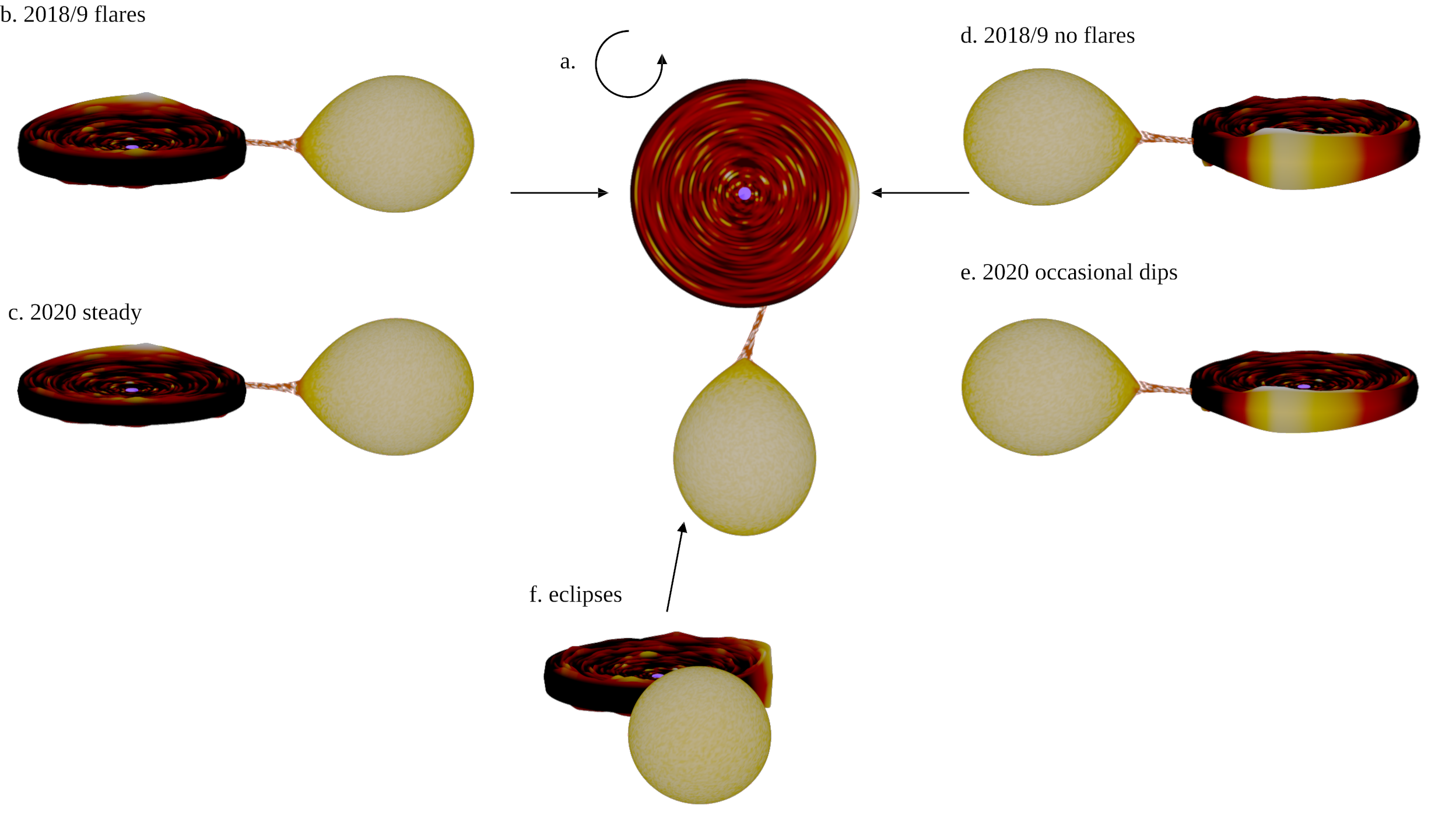}
\caption{Schematic diagram showing physical structure and cause of flaring behaviour in \seight. The binary system viewed down the orbital axis is shown in (a). The remaining panels show the system as observed at various orbital phases. Post-eclipse phase: during the flaring state, the primary source is often obscured but occasionally directly visible during flares (b); during the steady state, the primary source is always visible (c). Pre-eclipse phase: during the flaring state, the bulge always blocks the primary source (d); during the steady state, the bulge occasionally causes dips (e). The X-ray source is eclipsed by the donor regardless of accretion state (f).
The system is shown to scale ($1:10^{11}$) for $q=0.3$ ($i=77^\circ$), apart from the X-ray emitting region (purple), which has been increased for visibility. Images were produced with \textsc{binsim} \citep{hynes02}, modified to give a textured surface of the accretion disc.}
\label{fig:diag}
\end{figure*}

For the case of an extended disc atmosphere or wind, we can estimate a minimum size of the scattering material based on its being larger than the occulting region of the star. The scattered emission must be larger in at least one of radius (in the plane of the disc) and height (perpendicular to the disc). These constraints are shown in Figure~\ref{fig:scat}: the scattering material must have a height $>1.4\times10^8$\,m and/or a width $>10^9$\,m.
These sizes are of similar order to disc winds/atmospheres which have been observed or inferred in this \citep{munoz20,buisson20rgs} and other sources, which are generally a significant fraction of the disc \citep[e.g.][]{diaztrigo16}.

We also check whether a disc atmosphere could contain enough material to scatter a few per cent of the incident flux. If the radiation must pass through a distance $l=10^9$\,m of material of at least $n_{\rm e}=10^{17}$\,m$^{-3}$ \citep{buisson20rgs}, the fraction of radiation scattered is at least $nl\sigma_{\rm T}\approx10^{-2}$, so a few per cent being scattered is reasonable.

\subsection{Companion star}

We also find that the companion star is significantly larger than a main-sequence star of equal mass (Figure~\ref{fig:mr}). There are several possible reasons for its inflated size.
It may be due to the stellar evolution of the companion star itself, through reaching the end of core hydrogen burning. If this is the case, the star can only have very low mass if substantial mass transfer has already occurred to reduce its mass, since otherwise it would not have had time to finish core burning.
Alternatively, binary interactions, such as tides, irradiation or mass loss may have contributed to the inflation. In the case of irradiation, the star would need to fill its Roche lobe before the neutron star can accrete effectively to provide the source of radiation, unless the orbit was previously much smaller. The binary evolution tracks produced by \citet{podsiadlowski02} do not include the required period increases for relevant systems, so irradiation seems unlikely. Further to this, most of the systems considered by \citet{podsiadlowski02} that are in feasible regions of the mass-period plane have hydrogen free cores, so nuclear evolution is the most likely reason for the companion size.

\subsection{Physical scenario for flaring behaviour}

\seight\ is remarkable for the strong variability seen in its discovery state. The phase dependence of the flares suggests that variable obscuration from structure in the surface of the disc plays a significant role in enhancing the variability. A schematic diagram of this is shown in Figure~\ref{fig:diag}.

The dependence of flares on the orbital phase implies that they are governed, at least in part, by large scale structures of the system, since any phase information is rapidly lost closer to the neutron star, as the orbital time scale is much faster.
The most likely mechanism for the suppression is obscuration by the accreting material: the stream and disc. This implies that at phases prior to the eclipse, there is more material in our line of sight, due to a thicker structure.
These phases are where the interaction between the accretion stream and disc occurs: the stream drifts forward in phase from the Lagrange point (by a small amount) and as the material interacts with the disc, it is forced further forwards due to the higher (Keplerian) velocity of the disc material.
The interaction between these two components with different velocities is likely to produce heating and turbulence, which will thicken the resulting structure.
For this structure to suppress flares only at the phases observed, this thickened structure must settle again within around half an orbit.

Having explained the lack of flares at some phases by absorption from the surface of the accretion structure, we may consider whether a similar mechanism explains the flares themselves.
The apparent flares could represent the true intrinsic accretion flux, with the non-flare times being due to obscuration in the disc surface. 
This is broadly consistent with the changes in hardness seen in X-ray data: the bright flares are always relatively soft and show less absorption \citep{hare20}.
This model also requires structure in the thickness of the disc, such that a direct line of sight to the central region is occasionally, and only occasionally, present. Disc structure is also used to explain LMXB dippers \citep{frank87}; the ubiquitous variability of accreting sources \citep[e.g.][]{lyubarskii97,uttley01}; and is seen in numerical simulations of discs \citep[e.g.][]{hogg16}.
Therefore, it seems likely that variable absorption is responsible for much of the increase in variability amplitude compared to that seen in most accreting sources.

There must still be some variability in the intrinsic emission, since the radio is variable \citep{vandeneijnden20}, which comes from too large a region to be obscured by the disc surface. Additionally, the flux during the eclipse -- which must be scattered from a larger region than the companion star -- is also somewhat variable, probably from scattering a variable primary source.

\subsection{Similar sources}

Finally, we consider how \seight\ compares with other XRBs.

\exo\ shows intervals of strong dipping, which in soft X-rays can appear similar to the flaring behaviour of \seight, although \exo\ shows little simultaneous change at harder energies \citep{homan03}. This may be due to a similar variable obscurer, but that has a lower maximum column density than for \seight, so never fully obscures hard X-rays. Since \seight\ is a larger system, such a difference would not be unexpected.

Vela~X-1 is a high-mass XRB, which also shows variable emission close to the eclipse due to structure in the wind of the supergiant companion \citep{sidoli15}. It also shows an occasional secondary dip in the orbital light curve at inferior conjunction, attributed to scattering in an ionised accretion wake \citep{malacaria16}. These differences show that the different components in the accretion structure give different contributions to the full system when the system parameters are different.

Swift J1357.2--0933 shows quasi-periodic optical dips with an evolving period that is always much shorter than its orbit, implying that the dipping material is from closer to the centre of the disc. The reason for this difference -- either in terms of the waveband which shows dipping or the location of the obscuring material -- is not yet clear, but a connection with the presence of winds has been proposed \citep{corralsantana13,charles19,paice19,jimenezibarra19}.

Overall, these results show the power of observations of high inclination sources to elucidate the variety of accretion structure present in X-ray binaries.

\section*{Acknowledgements}

We thank the referee for helpful comments.
D.J.K.B. and D.A. are funded by the Royal Society.
T.M.D. and M.A.P. acknowledge support from the Sate Research Agency of the Spanish Ministry of Science, Innovation and Universities and the European Regional Development Fund under grant AYA2017-83216-P. T.M.D. acknowledges support via the Ram\'on y Cajal Fellowship RYC-2015-18148.
N.D. is supported by a Vidi grant from the Netherlands Organization for Scientific Research (NWO).
J.v.d.E. is supported by a Lee Hysan Junior Research Fellowship awarded by St. Hilda's College.
J. Hare acknowledges support from an appointment to the NASA Postdoctoral Program at the Goddard Space Flight Center, administered by the Universities Space Research Association under contract with NASA.
C.M. is supported by an appointment to the NASA Postdoctoral Program at the Marshall Space Flight Center, administered by the Universities Space Research Association under contract with NASA.
M.O.A. acknowledges support from the Royal Society through the Newton International Fellowship programme.
D.J.W. acknowledges support from STFC in the form of an Ernest Rutherford Fellowship.
\nicer\ is a mission of NASA's Astrophysics Explorers Program. This research has made use of data and software provided by the High Energy Astrophysics Science Archive Research Center (HEASARC), which is a service of the Astrophysics Science Division at NASA/GSFC and the High Energy Astrophysics Division of the Smithsonian Astrophysical Observatory.

\section*{Data availability}

The data underlying this article are available in HEASARC.

\bibliographystyle{mnras}
\bibliography{swiftj1858}

\section*{Affiliations}
\textit{\\
\noindent$^{1}$Department of Physics and Astronomy, University of Southampton, Highfield, Southampton, SO17 1BJ\\
  $^{2}$Instituto de Astrof\'isica de Canarias, 38205 La Laguna, Tenerife, Spain\\
  $^{3}$Departamento de Astrof\'\i{}sica, Universidad de La Laguna, E-38206 La Laguna, Tenerife, Spain\\
  $^{4}$NASA/Goddard Space Flight Center, Code 662, Greenbelt, MD 20771, USA\\
  $^{5}$Department of Astronomy, University of Maryland, College Park, MD 20742, USA\\
  $^{6}$Anton Pannekoek Institute for Astronomy, University of Amsterdam, Science Park 904, 1098 XH, Amsterdam, the Netherlands\\
  $^{7}$ESO, Karl-Schwarzschild-Strasse 2, D-85748 Garching bei M\"unchen, Germany\\
  $^{8}$ Department of Physics, Astrophysics, University of Oxford, Denys Wilkinson Building, Keble Road, Oxford OX1 3RH, UK\\
  $^{9}$Facultad de Ciencias Astron\'omicas y Geof\'{\i}sicas, Universidad Nacional de La Plata, Paseo del Bosque s/n, 1900 La Plata, Argentina\\
  $^{10}$Instituto Argentino de Radioastronom\'{\i}a (CCT-La Plata, CONICET; CICPBA), C.C. No. 5, 1894 Villa Elisa, Argentina\\  
  $^{11}$Eureka Scientific, Inc., 2452 Delmer Street, Oakland, CA 94602, USA\\
  $^{12}$SRON, Netherlands Institute for Space Research, Sorbonnelaan 2, 3584 CA Utrecht, The Netherlands\\  
  $^{13}$NASA Marshall Space Flight Center, NSSTC, 320 Sparkman Drive, Huntsville, AL 35805, USA\\
  $^{14}$Universities Space Research Association, Science and Technology Institute, 320 Sparkman Drive, Huntsville, AL 35805, USA\\  
  $^{15}$Kapteyn Astronomical Institute, University of Groningen, PO Box 800, NL-9700 AV Groningen, the Netherlands\\
  $^{16}$MIT Kavli Institute for Astrophysics and Space Research, Massachusetts Institute of Technology, Cambridge, MA 02139, USA\\
  $^{17}$Department of Astronomy and Space Sciences, Atat\"{u}rk University, Yakutiye, 25240 Erzurum, Turkey\\
  $^{18}$Astrophysics Science Division and Joint Space-Science Institute, NASA's Goddard Space Flight Center, Greenbelt, MD 20771, USA\\
  $^{19}$Department of Physics, Tor Vergata University of Rome, Via della Ricerca Scientifica 1, I-00133 Rome, Italy\\
  $^{20}$INAF - Astronomical Observatory of Rome, Via Frascati 33, I-00078 Monte Porzio Catone (Rome), Italy\\
  $^{21}$Space Sciences Laboratory, 7 Gauss Way, University of California, Berkeley, CA 94720-7450, USA\\
  $^{22}$Institute of Astronomy, Madingley Road, Cambridge, CB3 0HA\\
}
  
\bsp	
\label{lastpage}

\end{document}